\documentclass[doublecolumn]{elsart}

\usepackage{natbib, graphicx}

\usepackage{amssymb}

\journal{Journal of Informetrics}
\begin{document}

\begin{frontmatter}

\title{Communities, Knowledge Creation, and Information Diffusion}


\author[lambi]{R. Lambiotte},
\corauth[cor]{Corresponding author.} 
\ead{r.lambiotte@imperial.ac.uk}
\author[pietro]{P. Panzarasa}
\ead{p.panzarasa@qmul.ac.uk}

\address[lambi]{Institute for Mathematical Sciences, Imperial College London, Prince$'$s Gate 53, SW7 2PG London, UK}
\address[pietro]{School of Business and Management, Queen Mary College, University of London, Francis Bancroft Building, Mile End Road, E1 4NS London, UK}

\begin{abstract}
In this paper, we examine how patterns of scientific collaboration contribute to knowledge creation. Recent studies have shown that scientists can benefit from their position within collaborative networks by being able to receive more information of better quality in a timely fashion, and by presiding over communication between collaborators. Here we focus on the tendency of scientists to cluster into tightly-knit communities, and discuss the implications of this tendency for scientific performance. We begin by reviewing a new method for finding communities, and we then assess its benefits in terms of computation time and accuracy. While communities often serve as a taxonomic scheme to map knowledge domains, they also affect how successfully scientists engage in the creation of new knowledge. By drawing on the longstanding debate on the relative benefits of social cohesion and brokerage, we discuss the conditions that facilitate collaborations among scientists within or across communities. We show that successful scientific production occurs within communities when scientists have cohesive collaborations with others from the same knowledge domain, and across communities when scientists intermediate among otherwise disconnected collaborators from different knowledge domains. We also discuss the implications of communities for information diffusion, and show how traditional epidemiological approaches need to be refined to take knowledge heterogeneity into account and preserve the system's ability to promote creative processes of novel recombinations of ideas. \end{abstract}

\end{frontmatter}

\section{Introduction}

The recent development of online libraries and efficient search engines allows us to have a quantitative description of a number of scientific collaboration networks based on a large amount of scientific papers, with precise details about the identity of the authors, the subject of the papers (keyword analysis) as well as the relations between these papers (citations). This development offers exciting new perspectives and opportunities for understanding how the process of scientific production is organized and evolves over time. This requires not only the mapping of the intellectual contributions and the scientists that make them, but also the study of how information flows among scientists and how they interact with one another. Electronic databases enable us precisely to trace the way scientists exchange, discover and create new information over time, which may help uncover the conditions and mechanisms underpinning successful transfer and sharing of knowledge, scientific productivity and creativity, such as the development of new areas of investigations and research topics. One way to study how scientists exchange and share information is through the construction of co-authorship networks \citep{revN}. When analyzing these networks, one may reasonably assume that the authors collaborating on a paper know each other (at least in relatively small collaborations) and have put their expertise in common in order to carry out joint research and co-write the paper. Similarly, citation analysis \citep{garfield2, garfield1, Leydesdorff} is a tool for evaluating how the ideas and concepts of a paper are used in subsequent works, leading to cascades of influence. In both co-authorship and citation networks, scientific collaboration is typically described in terms of a very large network, usually composed of tens of thousands of nodes, thereby lending itself to statistical description and motivating an analysis that combines the social sciences with complex network theory.

Some of the statistical quantities typically used to describe these networks are purely local and may be employed in order to give a measure of the ÒqualityÓ of a paper depending on its topological properties. The best-known example is certainly the in-degree of a paper, which is the number of its citations, and represents a standard way for quantifying its impact \citep{garfield2, wuchty}. The corresponding global description is the degree distribution, which is well-known to have a long tail for a wide range of different networks \citep{barabasi}. For example, this tail is well fitted by a power-law function in the case of citation networks \citep{redner} and of co-authorship networks \citep{collaboration}. Other local measures of the topology of the networks include the clustering coefficient, correlations between the degrees of adjacent nodes, etc.

The previous quantities give information about the local properties of the network around nodes. However, they do not help uncover the highly clustered nature of scientific production, namely the fact that co-authorship networks and citation networks are made of several dense groups of nodes, also called communities, such that there are many links between nodes of the same community and only few links between nodes of different communities \citep{GN}. Such a modular structure is often associated with the high specialization needed to perform research, and with the emergence of disciplines, their own jargon, interests and techniques \citep{Whitley}. A thorough understanding of this modular structure is important as it helps uncover the organization of scientific production. In this paper, we will first take a structural viewpoint and discuss how a collaboration network can be partitioned into communities by looking at the ties connecting two or more scientists when they co-write a paper. While these communities represent groups of nodes connected through dense overlapping ties, they may also suggest a possible organization of the network into clusters of nodes that are homogeneous with respect to some non-relational attribute. In particular, when communities of connected scientists also represent the set of individuals working in the same scientific disciplines, they may be used as a taxonomic scheme to map knowledge domains \citep{borner2003, boyack, chen03, loet} and to track the changing frontiers of science. 

The partitioning of scientific collaboration networks into communities that overlap with the organization of the network into distinct disciplines or research areas has important implications in terms of the performance of the scientists working within or across communities. Research in the social science has long been concerned with this issue, and has been marked by a sharp debate between two apparently opposed views. One view stresses the benefits of ``closed'', dense, or cohesive networks for performance \citep{coleman}, while the other emphasizes the value derived from ``open'', sparse, or brokered networks, rich in structural holes \citep{burt, Granovetter1973}. We build on, and extend, this debate on the trade-off between social cohesion and brokerage by investigating the conditions under which scientists can undertake successful work by collaborating with others within or outside their own communities. Moreover, the partitioning of the network into communities may have important implications in terms of information diffusion, especially as a result of the sporadic interactions between nodes in different communities. A related well-known example is that of the synchronization of oscillators on a modular network, in which synchronization takes place very fast within modules, but at a much slower time scale at the global level \citep{syn1, syn2}. The presence of communities is also known to have a profound impact on the emergence and survival of cooperation \citep{cooperation}, and on the possibility for heterogeneous ideas to co-exist in the system \citep{lambi}. 

The goal of this paper is to study the role of communities in knowledge creation from different angles. We will first focus in section 2 on the methods that have been developed in order to uncover communities in large networks, and propose a method that allows us to study networks of unprecedented size. This section will take a structural perspective and will be dedicated to a description of the network topology. In section 3, we will extend our structural analysis of communities by discussing whether and the extent to which they facilitate scientific production. In section 4, we will focus on the diffusion of information and examine how communities affect the creation and spreading of new ideas. The last section is dedicated to a discussion and summary of our main findings.

\section{Community Detection}

The way we access, use and analyze scientific knowledge has radically changed in the last few years due to the availability of a large amount of research databases on the Web, providing us with accurate and complete information about the content of scientific papers, their authors and their relations. As more information on scientific production continues to grow, new tools are needed in order to extract and organize knowledge, in the same way as Google helps us to find our way on the Internet. There are several, often complementary, areas of investigation that require suitable methods of analysis. Among these areas are, for instance, the identification of major researchers or keystone articles \citep{chen07}, the discovery of new articles based on readers' previous search behavior and interests \citep{Kautz}, and the analysis of emerging trends and the relations between different disciplines. In general, the aim of all these areas of study is to offer readable maps of knowledge domains.
 
There are several ways to investigate the organization of scientific production. This can be done at the level of the papers themselves, by imposing a classification scheme, such as the PACS classifications in the physics literature, or by organizing databases in terms of the semantic similarity of their contents \citep{Landauer}. Another approach consists in representing scientific production in terms of a complex network, where different kinds of nodes (authors and articles) and different kinds of links (who writes with who, what cites what) are present. This method has the advantage of being flexible, as it does not require a centralized organization into PACS classifications, thereby allowing for tracking the self-organization of science and the emergence of fields before a new specific journal has been created or before it has been recognized as a new category. This flexibility has a cost, however, as such network representations are still very complex, and require careful analysis in order to decompose the multitude of nodes and links into meaningful modules, and highlight the underlying structures and the relationships between them.

This problem is not specific to the mapping of knowledge domains as it occurs for almost any complex system that can be represented as a network, e.g., friendship networks, metabolic networks, and food-web networks \citep{GN, revN}. In general, a way to extract information from these very complicated multi-dimensional systems consists in uncovering their ``community structure'' \citep{Roswall}, namely in dividing the network into groups such as most of the links are concentrated within the groups, while there are only few links between nodes of different groups. In other words, this approach consists in finding a meaningful partition of the network into communities or sub-units. This partition may then be used in order to produce a coarse-grained description of the full network, by assuming that the nodes belonging to the same community are equivalent, and by considering a higher-level meta-network where the nodes are now the communities. The resulting meta-network whose nodes are the communities may then be used to visualize the original network structure. The identification of these communities is therefore of crucial importance, especially because they may overlap with (often unknown) functional modules such as topics in information networks, disciplines in citation networks, or cyber-communities in online social networks.

\begin{figure}
\includegraphics[width=0.7\textwidth]{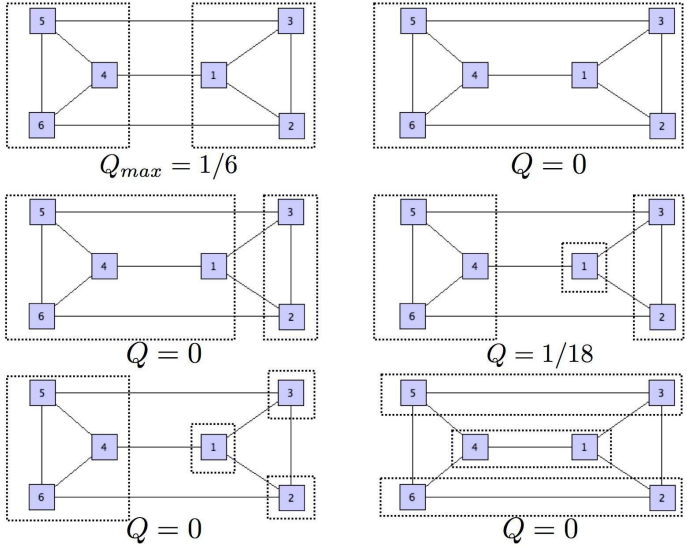}
\caption{Some of the partitions of a simple network made of 6 nodes and 9 links. The  partition with the highest modularity Q divides the system into 2 communities. In this case, the problem of finding the best partition is trivial, due to symmetry reasons, but it is much more complicated when the system is larger and less regular.}\label{fig1}
\end{figure}

In the last few years, there has been a concerted interdisciplinary effort to develop mathematical tools and computer algorithms to detect community structure in large networks \citep{N, NG, N06a}. Such a problem is often computationally intractable and therefore requires approximation methods in order to find reasonably good partitions in a reasonably fast way. The rapidity of the algorithm has become a crucial factor due to the increasing size of the networks to be analyzed. A large variety of methods have been developed in order to address this problem \citep{fort}. In this paper, we will focus on a type of approach which has proven particularly effective and which is based on the optimization of a quality function, the so-called modularity\footnote{The modularity of the partition of a network is given by  $Q = {1\over2m} \sum_{i,j} \biggl[ A_{ij} - {k_ik_j\over2m} \biggr]
    \delta(c_i,c_j)$, where the sum is performed over all pairs of nodes belonging to the same community, $m$ is the total number of links, $k_i$ the degree of node $i$ and $A$ is the adjacency matrix of the network. From a physics perspective, modularity can be interpreted as the Hamiltonian of a q-Potts model with nearest neighbours interactions \citep{reichardt}. 
    } \citep{NG}. The modularity of a partition is a scalar value between -1 and 1 that measures the density of links inside communities as compared to links between communities (see Figure 1). The exact optimization of modularity is a problem that is computationally hard \citep{BDG}. A number of algorithms have been recently introduced in order to deal with this problem. 
    For a comparison of the accuracy and computational cost of different methods, we refer to the excellent review by \citet{danon}. 
    The first method proposed to optimize modularity was the divisive algorithm of \citet{GN}. However, this method is very slow and has been outperformed by more recent methods \citep{N06b}.
    The best method in terms of accuracy is certainly Simulated Annealing; however, its applicability is limited to systems of relatively small size \citep{guimera}. Up to recently, the fastest algorithms were the greedy algorithm of \citet{CNM} and its generalization by \citet{WT}, which allowed researchers to analyze systems including up to a few million nodes.

In this paper, we will use a method which was introduced very recently and which outperforms previous methods in terms of computation time, while having an excellent accuracy \citep{blondel}. This method takes advantage of the self-similar nature of complex networks \citep{Havlin}, namely the fact that many networks observed in the real world are composed of several natural levels of organization, i.e., the networks are organized into communities that divide themselves into sub-communities \citep{arenasf,sales}. This Multi-Level Aggregation Method (that we call ``Louvain method" since now on) incorporates such a multi-level organization and consists of two phases that are repeated iteratively\footnote{For a detailed description of the Louvain method and its properties, we refer to the original paper by \citet{blondel}. C++ and matlab versions of the program are freely available at {\it http://findcommunities.googlepages.com} }. First, the algorithm looks for ``small'' communities by optimizing modularity in a greedy, local way. Second, the algorithm aggregates nodes of the same community and builds a new network whose nodes are the communities. These phases are repeated iteratively until a maximum of modularity is attained and an optimal partition of the network into communities is found. The choice of this community detection method is motivated by its excellent accuracy and its rapidity which allows to uncover networks of unprecedented size (for example, in \citet{blondel}, a network of more than 100 million nodes is analyzed in around 2 hours). The rapidity of the algorithm therefore opens exciting opportunities, as it allows us to analyze networks made of millions of nodes, and therefore to study whole datasets, instead of dividing them into sub-parts due to limitations of computation time. This rapidity also enables us to study the evolution of large networks (and therefore the birth, death, merging, etc. of communities), by focusing on several snapshots taken at different points in time \citep{evolution}.

We now apply this algorithm to the co-authorship network of the scientists that posted preprints on the Condensed Matter E-Print Archive. To construct the network, we have included all preprints posted between Jan 1, 1995 and March 31, 2005. This network, whose statistical properties have been described in \citet{collaboration}, exhibits typical features of social networks, such as a high clustering coefficient and a fat-tailed degree distribution. It is composed of N= 40421 scientists and of L=175693 links. The Louvain method finds a partition of modularity $Q=0.729$ (made of 1032 communities) in less than 1 second. For the sake of comparison, the method of \citet{CNM} finds a worse partition of modularity $Q=0.654$ in more than 4 minutes. It is also interesting to note that the difference in accuracy and in computation time decreases for a random network where the links between the nodes have been randomly redistributed. In this case, it takes 8 seconds to the Louvain method to find a modularity of $Q=0.283$ (as expected, this value of modularity is smaller than in the case of the original network), while the method by \citet{CNM} finds a modularity of $Q=0.277$ in 80 seconds. The fact that the Louvain method is slower for a random network arises from the absence of internal structure in the random network, which makes the multi-level approach less efficient. It is interesting to note, however, that also in the case of a random network the Louvain method is still more rapid and accurate than the alternative method. One should note that the Louvain method has been recently applied to co-citation networks \citep{wallace} where it was shown that the uncovered communities correspond to coherent groups of research and are indeed representative of the structure of a given scientific discipline.

\begin{figure}
\includegraphics[width=1.0\textwidth]{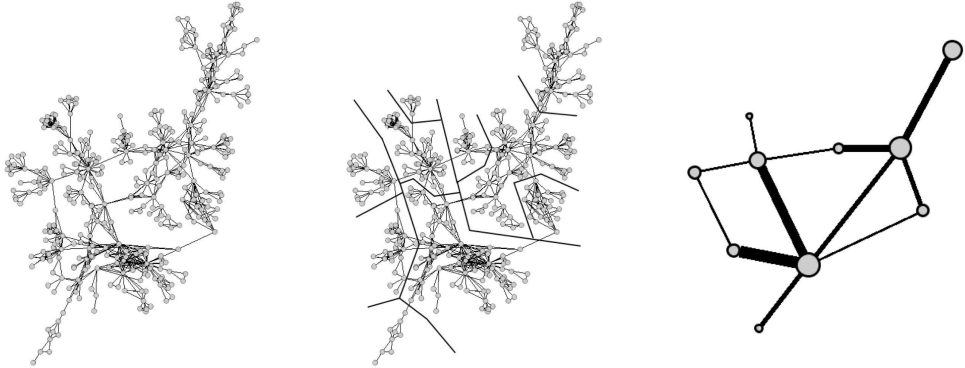}
\caption{By optimising modularity, one uncovers sets of topologically equivalent nodes and the relations between them, thereby allowing to represent the network in a coarse-grained manner.}\label{fig1}
\end{figure}

The visualization of the co-authorship network by using standard programs such as Visone or Pajek would not be very helpful, as the network would resemble a cloud with too many links and nodes to be distinguished. By contrast, by agglomerating nodes into communities, with an obvious reduction of the size of the system (from 40000 nodes to 1000 communities in the above example), and by highlighting the relations between these groups of nodes, the community detection method makes such a visualization possible. Let us illustrate this coarse-graining process by focusing on a smaller collaboration network of scientists working on network theory and experiment, which has been studied in detail in \citet{N06a}. This network is made of 
1589 scientists,  379 of which belong to the largest component. As shown in figure 2, the Louvain method partitions this largest component into 10 communities and allows to clarify the network representation.

\section{Social Structure and Knowledge Creation}

In the previous section, we have taken a structural perspective, and have shown how a scientific co-authorship network can be portioned into communities of tightly-knit scientists by looking at the links among scientists. In this section, we will explore the implications of network structure and its partition into communities in terms of the performance of the scientists. More generally, by having an impact on the degree to which nodes are exposed to the information flowing in a network, structure affects how successfully nodes undertake their tasks \citep{Smith-Doerr}. Among these tasks, we will concentrate on knowledge creation and scientific production, that here we broadly define to include all creative intuitions and combinatorial processes leading to scientific and technological advances through novel rearrangements of ideas, theories, methods, processes, strategies, and so on \citep{Burt2004, Fleming}.

The network foundations of knowledge creation have long been documented in the social sciences \citep{Allen, tushman}. Recent empirical studies have uncovered the positive effects of multi-authorship on research performance, suggesting that teams typically produce more frequently cited research than individuals do \citep{wuchty}. Moreover, researchers have been concerned with the mechanisms that underpin the influence of collaborative structures on human creativity not only within the domain of scientific endeavors, but also in the context of artistic production. For example, \citet{Uzzi2005} focused on a network of creative artists who made Broadway musicals, and found a nonlinear association between the ``small-world'' properties\footnote{``Small-world'' networks are built from a regular lattice where a fraction of the links is replaced by random links. By changing this fraction, one interpolates between a regular lattice and a random network \citep{Watts}. Such a model exhibits a high density of triangles as well as a small diameter.} of the collaborative network and the production of financially and artistically successful shows. In particular, they showed that when the clustering coefficient ratio is low or high, the financial and artistic success of the shows is low, while an intermediate level of clustering is associated with successful shows. 

While uncovering the ``small-world'' network effects on creativity, these results help shed light on a fundamental network mechanism that has long been investigated in the social sciences: social cohesion. Building on Coleman's (1988) conception of social capital, scholars have studied the benefits of cohesive social structures organized into well-defined tightly knit communities of connected individuals. In particular, the tendency of individuals to forge links locally within groups has often been associated with an increase in one's social capital, in that it engenders a sense of belonging, fosters trust, facilitates the enforcement of social norms, and enables the creation of a common culture \citep{Reagans2003, Uzzi1997, Uzzi2005}. For example, if individual $A$ has links with individuals $B$ and $C$, a link between $B$ and $C$ would enable the three individuals to detect and punish one another's undesirable behavior more easily, increasing the expected costs of opportunistic behavior with respect to the case in which a link between $B$ and $C$ is absent. Mutual monitoring abilities will in turn engender trust among connected individuals and sustain the generation of group norms more easily and to a greater extent than would be the case if individuals did not have dense and overlapping links with one another. 

By fostering trust and promoting the enforcement of social norms, social cohesion that occurs within communities offers the facilitating conditions for coordination and collaborative endeavors. For example, an abundance of empirical evidence supports the idea that links embedded in social relationships reduce competition and increase the motivation to transfer information. If people who trust one another are more likely to exchange information, cohesion will then enhance information sharing. Individuals in cohesive communities will be able to obtain information in a timely fashion, and will also benefit from the exchange of complex, tacit and proprietary information \citep{Hansen, Uzzi1997}. More complete information that can be obtained more easily will in turn facilitate innovation and knowledge creation \citep{Ahuja, Obstfeld}. Moreover, by engendering a supportive social context, trust sustains risk-taking and learning, with further positive effects on scientific creative endeavors \citep{Amabile, Edmonson}. 

Despite all the benefits associated with social cohesion, nonetheless the tendency of individuals to cluster into tightly knit communities also bears a cost: local redundancy. From a dynamic perspective, the more an individual's additional contacts are already connected to the individual's current ones, the less likely they are to take the individual closer to new people he or she does not know already. Lack of connections with new social circles may create isolation and eventually degrade performance. Building on the seminal arguments of \citet{Granovetter1973} and \citet{burt}, proponents of the benefits of brokerage point out that in cohesive networks organized into communities links tend to be strong as people invest a disproportionately large amount of their time and resources in relationships with few others. Cohesive networks thus make links with dissimilar others and exposure to new information less likely. By contrast, in networks that are rich in structural holes, where individuals broker between otherwise disconnected contacts, links tend to be weak and more likely to connect people with different ideas, interests and perspectives \citep{Burt2004}. If scientific production requires prompt access to novel information, then people embedded in brokered structures will be more creative and successful in their endeavors. From this perspective, brokers between communities occupy the most advantageous boundary position as they lie at the confluence of fresh and heterogeneous ideas that they can creatively integrate into novel recombinations \citep{Brass, Burt2004}.

While scholars in the social sciences agree on the importance of social structure for information diffusion and performance, there is still controversy over the optimal structure and, more specifically, over the relative benefits of social cohesion within communities on the one hand, and brokerage between communities on the other. A number of empirical studies have suggested that an appropriate combination of density and sparseness can provide individuals with the necessary redundant trusted relationships and access to non-redundant external contacts that will enable them to successfully perform their tasks \citep{Burt2005, Podolny}. A more recent line of investigation has examined the apparent trade-off between social cohesion and brokerage by focusing on the interactions between network structure and the attributes of the interacting individuals \citep{Perry-Smith, Reagans, Rodan}. For example, \citet{Fleming} have empirically examined the mitigating effects exerted by individuals' attributes on the benefits associated with brokerage. Their study suggests that, while brokerage between otherwise disconnected collaborators makes all individuals more likely to create new ideas, at the same there are marginal contingent positive effects of social cohesion on generative creativity when individuals and their collaborators bring broad experience, have worked for multiple organizations, and have connections with external contacts. 

A combined study of network structure and nodes' attributes becomes especially relevant in the context of knowledge creation, scientific production and innovation, where the benefits of social relationships crucially depend on the information scientists already possess as well as on the heterogeneity and breadth of the information they can obtain from their contacts. More generally, there is little consensus on the effects of access to heterogeneous knowledge on performance. On the one hand, recent work has examined the benefits that scientists can gain from specializing, in terms of research productivity, promotion, tenure standards and academic earnings \citep{Leahey}. On the other, there is also evidence that access to novel heterogeneous information is beneficial for creativity and innovation \citep{Burt2004, Hargadon}. For example, \citet{Rodan} have found that the variety of knowledge to which managers are exposed positively affects not only their overall performance, but also their ability to accomplish complex tasks, create and implement novel ideas. 

With only few exceptions \citep{Guimera2005}, and despite the importance of knowledge heterogeneity and inter-disciplinarity for scientific production, scanty attention has been devoted to the way collaborative structures combine with scientists' degree of specialization and access to pools of diverse knowledge to affect their research performance. Scientists can vary the breadth of access to knowledge by carefully building their networks and selecting their collaborators either within their own specialty area or in different areas that enable them to obtain knowledge without having to acquire it personally. On the one hand, scientists can reduce access to heterogeneous knowledge by selecting their collaborators within their own specialty area. In so doing, they enhance scientific consensus, and facilitate scientific production through the generation of shared norms of research practice \citep{Moody}. On the other, scientists can expand access to heterogeneous knowledge by engaging in collaborations with other scientists from different specialty areas. While scientists typically rely on their collaborators to obtain the knowledge and expertise they do not have already \citep{Laband}, research has largely overlooked the various collaboration patterns through which scientists control their access to heterogeneous knowledge pools, and how these patterns ultimately affect knowledge creation and research performance.

Recent empirical work has investigated the extent to which the interplay between knowledge heterogeneity and the structure of the collaboration network affects a scientist's ability to produce research of high impact \citep{Panzarasa}. Drawing on the collaboration network of the social scientists that authored or coauthored the publications submitted to the 2001 Research Assessment Exercise in business and management in the UK, this work shows that scientists bridging two otherwise disconnected contacts with heterogeneous knowledge have a better performance than scientists with no such brokerage opportunities. At the same time, highly cited scientists also tend to be socially embedded within communities in which knowledge is homogeneously distributed across members. In this case, when scientists and their collaborators are not diverse in knowledge, collaborations are beneficial when they occur with contacts that are already collaborating themselves.

This work adds a new dimension to the relevance of communities for knowledge creation and scientific performance, and more generally to the debate on the relative benefits of social cohesion and brokerage. On the one hand, when scientists seek collaborators within their own knowledge pool, they can enhance their research performance by generating structurally cohesive communities. Thus, while communities often serve as a taxonomic scheme to map knowledge domains \citep{borner2003, boyack, chen03}, they also offer the supportive structural conditions for the successful performance of collaborative scientific work that remains confined within the boundaries of a knowledge domain. On the other, when collaborative scientific work spans across knowledge domains, scientific performance increases when scientists intermediate between their collaborators. Bridging structural holes between otherwise disconnected knowledge pools creates linkages across distinct scientific communities that offer knowledge brokerage opportunities for novel recombination of ideas. 

\section{Information diffusion and knowledge heterogeneity}

From a modeling point of view, innovation and knowledge creation can be seen as a catalytic process \citep{andrea,andrea2,Hanel,Lambiotte}. The juxtaposition of ideas in the mind of an individual may lead to syntheses and to the emergence of new ideas that can then diffuse and reach other individuals and cascade through the social network. This propagation may in turn result in further syntheses and in the emergence of other new ideas which are then diffused and so on, thereby leading to a sequence of self-reproducing flows of new ideas. In principle, a good model for innovation and knowledge creation should therefore incorporate these two types of ingredients: synthesis and diffusion. Diffusion has been studied extensively \citep{Bettencourt, Goffman1964, Goffman1966,rogers}, especially because of its parallel with the dynamics of an epidemic. Like a disease and its propagation, a new idea typically spreads among people that communicate directly (e.g., by talking, or via telephone and e-mail) or indirectly (e.g., by reading the same journals) \footnote{In this paper, we focus on models where a process diffuses on a static network. This limitation can be overcome by looking at the co-evolution of diffusion and of network dynamics \citep{schweitzer,castel}.}. This parallel has motivated the modeling of the evolution of scientific fields as epidemiological contact processes such as the Susceptible-Infected-Recovered (SIR) model or the discrete-time Independent Cascade Models\footnote{In the Independent Cascade Model, one starts from an initial set of infected nodes. When a new node becomes infected, it tries one single time to infect each of its neighbors with independent probability $p$. The process stops when no new node has been infected and is available to continue the propagation.} \citep{goldenberg2001tnc, kempe2003msi}.

Mathematical epidemiologists have long emphasized the important role played by the network topology in determining properties of disease invasion, spread and persistence \citep{May}. Several general results have been derived, such as the fact that epidemics spread without a threshold on a scale-free network due to the presence of hubs\footnote{In the context of the diffusion of innovations, the importance of the heterogeneity of the agents is well-known and usually taken into account by categorizing them into categories, e.g. innovators, early adopters, etc. \citep{rogers} or introducing opinion leaders \citep{valente}.}, which accelerate the diffusion by reaching an unusually high proportion of other nodes \citep{pastor}. Another important result is that diffusion is more efficient in random networks than in clustered networks, and that the presence of random links is fundamental for promoting diffusion \citep{Huang, vazquez}. This result, which supports the Granovetter's (1973) hypothesis of the strength of weak ties, is due to the fact that random links minimize the accumulation of several contacts around the same nodes, thereby reducing redundant links and accelerating diffusion across different parts of the network. 

This result, however, needs to be critically re-assessed in the light of our previous discussion about social cohesion. In section III, we noted that cohesive structures are likely to foster trust and facilitate knowledge transfer and sharing. Unlike the above mentioned results on disease spread, our analysis thus suggested that dense networks clustered into communities may accelerate, at the very least locally, information diffusion. This observation, therefore, cautions against a direct application of epidemiological models to a knowledge diffusion context. In order to explore this issue, researchers have recently modified the above models of disease spread in order to preserve and enhance the role of social cohesion \citep{watts2002smg}. For instance, threshold models are based on the fact that infection requires simultaneous exposure to multiple active neighbors \citep{granovetter1978tmc,kempe2003msi}. Similarly, generalized cascade models are based on the fact that the probability for a node to get ``activated'' depends on the number of times it has been in contact with an idea \citep{dodds2004ubg, kleinberg:cbn}. Within the context of ``small-world'' networks, research has shown that different types of links (random vs. regular) play very different roles in the propagation of ideas \citep{p1p2}. Random links, which are short-cuts in the network connecting otherwise distant regions, play an integrative role by connecting different communities, and therefore enable nodes to be exposed to, and explore, different parts of the network. Regular links, i.e. links connecting neighbouring nodes, by contrast, connect nodes within communities, and are likely to increase the number of infected paths available to each node. More interestingly, it was also shown that, when redundancy is needed to secure infection and adoption of a new idea, the presence of random links may actually hinder the emergence of cascades, and that the size of the avalanche depends in a non-trivial way on the modular structure and on the model parameters \citep{centola, centola2, p1p2}.

This epidemiological approach typically focuses on the diffusion of one idea. Starting from one ``infected'' individual, researchers are interested in the way an idea propagates among acquaintances in the social network, and try to estimate the total number of ``infected'' individuals. This approach, however, neglects the catalytic nature of knowledge creation, namely the fact that several ideas diffuse in the system and at the same time may be creatively recombined to produce new ideas. More specifically, what is often ignored is the role played by heterogeneity between ideas, a property that most epidemiological models forget to take into account by simply assuming homogeneity throughout the system. It is interesting to note that the catalytic nature of knowledge creation calls for a critical reassessment of the network implications for information diffusion. On the one hand, a rapid diffusion of ideas is crucial as it facilitates knowledge creation by increasing the ideas that individuals can obtain and recombine. On the other, however, if ideas reach too many people too quickly, they might generate consensus and lead to convergence toward a popular, though smaller, set of shared ideas, thereby hindering the capacity of innovation of the system \citep{Fang}. In this sense, the presence of modules, or niches, is necessary in order to ensure the co-existence of several ideas and preserve the fundamental diversity of knowledge conducive toward the production of further new knowledge. This observation has found support, for instance, in the context of opinion dynamics, where the fragility of consensus under variations of the network topology was highlighted \citep{lambi}. 

\section{Conclusions and discussion}

In this paper, we have focused on the role played by communities in knowledge creation. By integrating approaches from graph theory, economics, sociology and physics, we highlighted the relations between network structure, performance, and information diffusion, in the specific context of scientific production. In section II, we introduced the concept of community at the network level, by focusing on links between authors, regardless of the nodes' attributes. It was argued that the uncovering of communities is necessary in order to highlight relations between elements, reduce the dimension of the system and provide useful maps of knowledge. In section III, we explained how and to what extent communities can be advantageous for scientific performance. Since scientists can benefit from cohesive collaborations when their collaborators belong to the same knowledge domain \citep{Panzarasa}, communities at the network level will support scientific performance when they reflect unique non-overlapping knowledge domains. In this case, successful science production will therefore be organized into a disproportionately large number of cohesive collaborations among scientists with homogeneous knowledge within the same community, and relatively few brokered collaborations among scientists with heterogeneous knowledge across different communities. 
The interplay betwen communities and knowledge creation was then discussed from a modelling point of view in section IV, where we showed that the creation and diffusion of knowledge may be driven by different network mechanisms. On the one hand, random links facilitate rapid communication of ideas within the network. On the other, when redundancy is needed for individuals to adopt a new idea, the presence of local structure and communities not only accelerates diffusion due to the presence of redundant cohesive relationships, but also promotes diversity of knowledge across communities, thereby supporting the capacity of the system to innovate through creative recombinations of different ideas.

{\bf Acknowledgements} We would like to thank JL Guillaume for performing the analysis of the network of co-authorships with the method of \citet{CNM}.

\end{document}